\documentclass[acmsmall]{acmart}
\AtBeginDocument{%
  }

\copyrightyear{2025}
\acmYear{2025}
\setcopyright{acmlicensed}\acmConference[PEARC '25]{Practice and Experience in Advanced Research Computing}{July 20--24, 2025}{Columbus, OH, USA}
\acmBooktitle{Practice and Experience in Advanced Research Computing (PEARC '25), July 20--24, 2025, Columbus, OH, USA}
\acmDOI{10.1145/3708035.3736060}
\acmISBN{979-8-4007-1398-9/2025/07}

\begin{document}

\title{The National Research Platform: Stretched, Multi-Tenant, Scientific Kubernetes Cluster}

\author{Derek Weitzel}
\email{dweitzel@unl.edu}
\orcid{0000-0002-8115-7573}
\author{Ashton Graves}
\email{agraves10@unl.edu}
\orcid{0009-0002-9562-2621}
\author{Sam Albin}
\email{sam.albin@unl.edu}
\orcid{0009-0005-9688-1301}
\author{Huijun Zhu}
\email{huijun.zhu@unl.edu}
\orcid{0009-0005-8784-7224}
\affiliation{%
  \institution{University of Nebraska-Lincoln}
  \city{Lincoln}
  \state{Nebraska}
  \country{USA}
}

\author{Frank Würthwein}
\email{fkw@ucsd.edu}
\orcid{0000-0001-5912-6124}
\author{Mahidhar Tatineni}
\email{mahidhar@sdsc.edu}
\author{Dmitry Mishin}
\email{dmishin@ucsd.edu}
\orcid{0000-0003-1125-448X}
\author{John Graham}
\email{jjgraham@ucsd.edu}
\orcid{0000-0002-2139-5617}
\author{Elham E Khoda}
\email{ekhoda@ucsd.edu}
\orcid{0000-0001-8720-6615}
\author{Mohammad Firas Sada}
\email{mfsada@ucsd.edu}
\orcid{0009-0006-6045-2940}
\affiliation{%
  \institution{San Diego Supercomputer Center}
  \city{La Jolla}
  \state{CA}
  \country{USA}
}

\author{Larry Smarr}
\email{lsmarr@ucsd.edu}
\orcid{0000-0003-2003-7700}
\author{Thomas DeFanti}
\email{tdefanti@ucsd.edu}
\orcid{0000-0002-7642-2336}
\affiliation{%
  \institution{University of California, San Diego}
  \city{La Jolla}
  \state{CA}
  \country{USA}
}

\renewcommand{\shortauthors}{Weitzel et al.}

\begin{abstract}

The National Research Platform (NRP) represents a distributed, multi-tenant Kubernetes-based cyberinfrastructure designed to facilitate collaborative scientific computing. Spanning over 75 locations in the U.S. and internationally, the NRP uniquely integrates varied computational resources, ranging from single nodes to extensive GPU and CPU clusters, to support diverse research workloads including advanced AI and machine learning tasks. It emphasizes flexibility through user-friendly interfaces such as JupyterHub and low level control of resources through direct Kubernetes interaction. Critical operational insights are discussed, including security enhancements using Kubernetes-integrated threat detection, extensive monitoring, and comprehensive accounting systems. This paper highlights the NRP’s growing importance and scalability in addressing the increasing demands for distributed scientific computational resources.

\end{abstract}


\begin{teaserfigure}
  \centering
  \includegraphics[width=0.7\textwidth]{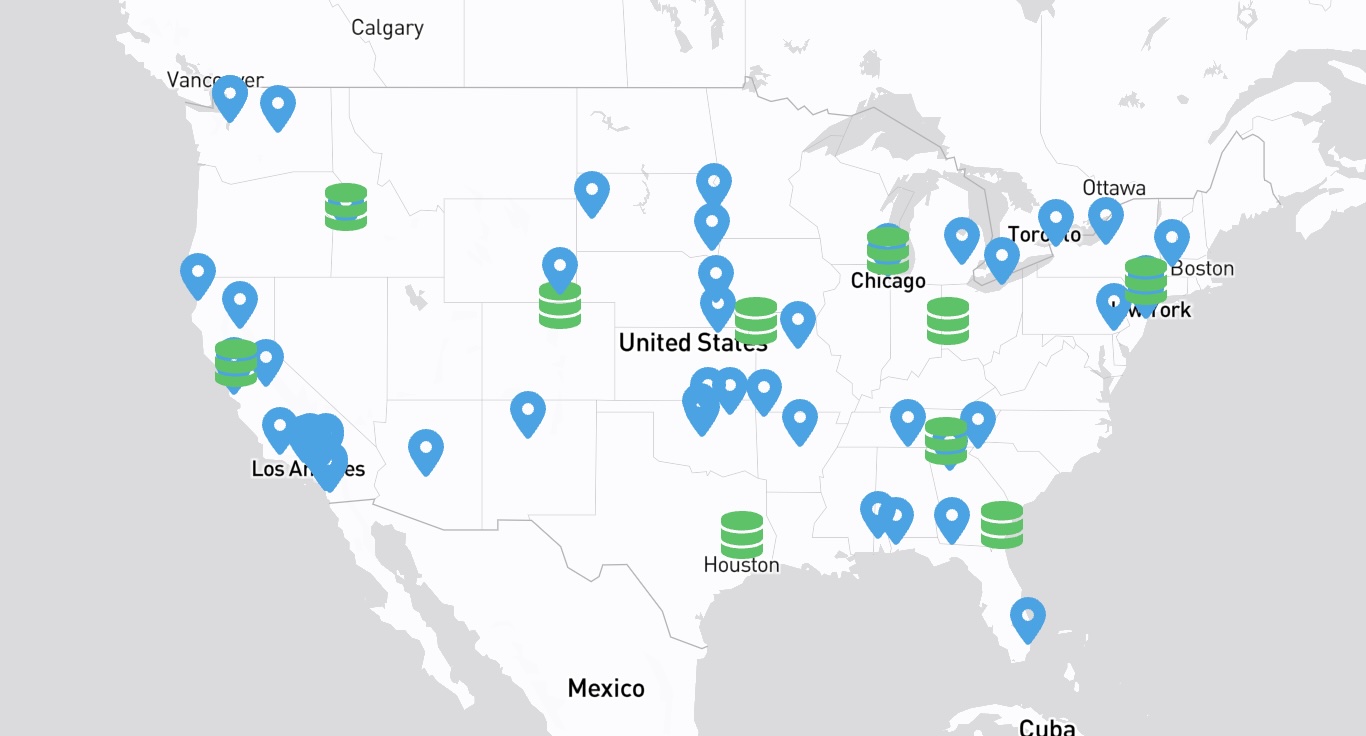}
  \caption{Map of NRP sites throughout the US.  Not pictured are the several international locations.}
  \Description{Map of NRP sites throughout the US.  Not pictured are the several international locations.}
  \label{fig:nrp-us}
\end{teaserfigure}


\maketitle

\section{Introduction}

The NSF supported research computing community has established infrastructures tailored to different levels of contributions.
ACCESS \cite{boerner_access_2023} provides an umbrella organization primarily for large resources, each of which are located in a single data center and run their own batch schedulers.
The OSG \cite{pordes_open_2007} federates clusters in different locations worldwide that each have their own batch scheduler. The OSG then overlays a fabric of services 
on top to federate the individual clusters to enable distributed High Throughput Computing\cite{raman2000resource}. The National Research Platform (NRP) federates individual compute and storage nodes into a global scale resource cluster across an a priori unlimited number of locations. The NRP thus operates at levels in the vertical software stack below either OSG or ACCESS. The entirety of the NRP, encompassing all globally distributed hardware locations, currently presents as a single `batch cluster' to users.


The NRP has attracted a growing number of users.  In calendar year 2024, there were 670 NRP Nautilus Namespaces that used ~5M GPU-hours and ~50M CPU Core-Hours on the NRP. These Namespaces can be divided into 4 main categories: Community Software (2), Observatories (13), NRP Operations (130), Individual campus researchers (525). These latter 525 Namespaces consumed >3/4 of the GPU-Hrs and 2/3 of the CPU-hrs used by all 670 NRP Nautilus Namespaces active in 2024. The individual PIs of these 525 researcher Namespaces came from 50 campuses, including 23 Minority Serving Institutions. These campuses include ones located in a dozen NSF designated Established Program to Stimulate Competitive Research (EPSCoR) \cite{epscor} States. The campuses span from Carnegie R1/R2 campuses to community colleges. This broad national breadth of campuses are critical to the future development of a US trained AI/ML workforce.

The NRP can federate resources ranging from a single node to large collections of resources located in the same data center, enabling groups to contribute hardware according to their capacity.
NRP offers integration at either the IPMI or the Kubernetes\cite{kubernetes} layer, thus releasing the resource owners of the burden of system and cluster administration, in addition to offering higher level services. The National Research Platform (NRP) originated from the Pacific Research Platform \cite{smarr_pacific_2018}, which began as a testbed for evaluating network connectivity across multiple sites. It has since evolved into a specialized platform focused on GPU hosting for Artificial Intelligence (AI) workloads. Additionally, the NRP provides educational support through fully-managed Jupyter notebook environments, LLM as a service, training resources, and curated containers. It also fosters community formation through a chat platform implemented via Matrix support channels.



Each institution contributes hardware to the NRP for its own reasons. The reduced burden of system administration attracts both institutions with limited resources as well as institutions for which research computing is not a primary focus (e.g. regional networking organizations and Internet2). Other institutions are attracted by the higher level services and the community building aspects of NRP. For example, institutions like the University of Missouri participate because their researchers already utilize the NRP and wish to expand their usage, including experimenting with new and/or unique hardware configurations \cite{hurt_adventures_2024}. For these researchers, the capability to run workloads on their dedicated nodes while seamlessly extending to additional research computing resources significantly amplifies their scientific impact. In addition, a number of projects use NRP as a distributed service deployment platform to build higher level services on top of NRP.


Some sites contribute hardware to meet funding requirements, such as the 20\% resource-sharing mandate in recent Campus Cyberinfrastructure (CC*) \cite{ccstar} solicitations. In these cases, the contributed hardware is shared among NRP users, and in return, the NRP user-support team actively assists these sites by training their local users. Additionally, the NRP systems team supports the site’s local management of the hardware and ensures compliance with any required institutional policies, monitoring security, and keeping software at all systems levels up-to-date. Increasingly, we see the community building aspects of NRP to lead to one institution hosting equipment bought by another institution, thereby contributing to the NRP community owned cluster by providing rackspace, power, and networking.


The continued growth in the NRP user community lead to the implementation of a richer set of policies. This includes a resource allocation mechanism for the most heavily in demand GPUs, an Acceptable Use Policy, and enhancements to security, monitoring, and their automation, as detailed in Section \ref{sec:security}. Overall, the NRP continues to grow in both users and resources, while providing innovative offerings such as hosted LLMs\footnote{Including those that provided assistance in editing this paper!} and other shared services.


The remainder of this paper describes the operations of the NRP, a distributed, multi-tenant Kubernetes cluster designed for scientific computing, and shares our experiences in supporting users to effectively leverage the platform.

\section{Distributed Deployment}

The Network Research Platform consists of over 420 nodes distributed across more than 70 sites as shown in Figure \ref{fig:nrp-us}. These sites include R1 institutions such as the University of California, San Diego, R2 institutions like the University of South Dakota, and network points-of-presence (PoPs) that provide high-quality network connectivity.

Managing and deploying the NRP management software across multiple administrative domains poses significant challenges. To overcome these hurdles, we utilize Ansible~\cite{ansible}, a powerful automation platform, to streamline the deployment and maintenance of our services. Specifically, Ansible simplifies the installation process of new nodes by automatically configuring and installing necessary software, including:


\begin{itemize}
    \item \textbf{Containerization}: The NRP installs the container runtime which will run the user pods.
    \item \textbf{Kubernetes}: The orchestration layer that the NRP uses to schedule workflows.
    \item \textbf{System settings}: The NRP sets numerous settings on the system such as increasing the default MTU on the network interface to support jumbo frames and turning on automatic security package updates.
\end{itemize}

Integration of resources into the NRP can happen at many different layers.

\begin{enumerate}
    \item \textbf{Hardware-Level Integration} - NRP administrators directly manage the node hardware using the Intelligent Platform Management Interface (IPMI), including installing and updating firmware and operating systems. For security, sites must restrict IPMI access exclusively to the designated NRP gateway through firewall policies.
    \item \textbf{Operating System-Level Integration} - Sites provide nodes pre-installed with NRP’s preferred operating system (Ubuntu). NRP administrators then install the necessary software with Ansible, integrate the nodes into the cluster, and handle ongoing OS maintenance and updates.
    \item \textbf{Peering Kubernetes} - In special cases, the NRP integrates resources by establishing peer relationships with external Kubernetes clusters using Admiralty\cite{admirality}. This allows secure, seamless sharing and processing of workloads across clusters.
\end{enumerate}

In order to keep track of all of the nodes in the NRP, we use NetBox\cite{netbox} as the source of truth for tracking hardware and network devices. It maintains an up-to-date inventory of nodes, interfaces, and other critical infrastructure. To automate the process of adding hardware, we deploy a customized version of the netbox-agent \cite{netbox-agent} on each NRP node. Additionally, NetBox’s journal feature is integrated with our Ansible workflows to log hardware changes and maintenance events, providing a transparent and auditable history of all actions performed on the infrastructure.

\section{User Interface}

Users primarily interact with the NRP either directly via Kubernetes or through gateways such as JupyterHub \cite{granger_jupyter_2021} and Coder \cite{coder}. To gain Kubernetes access, users log into the NRP portal to download a personalized Kubernetes configuration file containing Kubernetes RBAC-compliant OIDC tokens issued via CILogon federated authentication. Users are restricted to Kubernetes namespaces specific to their research groups, limiting their actions accordingly to within their namespace. Direct Kubernetes API interaction through command-line tools (e.g., \texttt{kubectl}) enables fine-grained resource management.

For interactive usage, the NRP hosts centralized instances of JupyterHub and Coder. The public JupyterHub instance simplifies container management through predefined hardware profiles, allowing users to easily select CPU, memory, and accelerator configurations while automatically provisioning persistent storage. It supports various software environments, including MATLAB, R, Python, X11 desktops, PyTorch, and TensorFlow. Similarly, Coder provides users with browser-based GUI Desktop/VS Code environments, complete with pre-configured persistent storage and templates for FPGA/GPU tools and libraries. Additionally, NRP documentation guides research groups in deploying customized JupyterHub and Coder instances within dedicated namespaces, while a hosted GitLab instance facilitates CI/CD workflows and Docker image management.


\subsection{Computational Resources}

As of Spring 2025, the NRP comprises in excess of 1,400 GPUs, 28,000 CPUs, and 161 TB of RAM distributed across more than 420 nodes. The GPU models available vary considerably, ranging from the older NVIDIA RTX 2080 Ti to the state-of-the-art Grace Hopper H200, with numerous intermediate models. Users can specify preferred GPU models or provide a list of acceptable GPUs. The Kubernetes scheduler then allocates resources accordingly.

\begin{figure*}[ht]
\centering
\includegraphics[width=\textwidth]{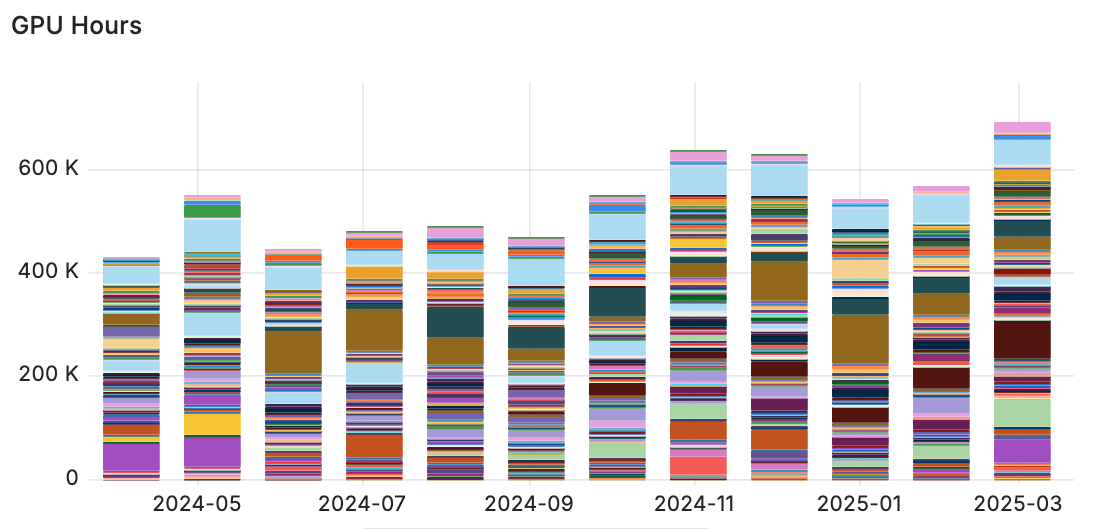}
\caption{GPU hours by research group on the NRP over time showing growth of GPU use. For the 12 month period, the total is 6,605,042 GPU hours from 412 research groups.}
\Description{A bar graph showing the GPU usage by month of the NRP for the previous year from March 2024 to March 2025. The graph shows the growth in usage of GPUs in the NRP from roughly 400k hours a month to over 600k hours, totaling 6,605,042 hours over the 12 month period from 412 research groups.}
\label{fig:gpuusage}
\end{figure*}

With the rising demand for high-end GPUs, such as the growing demand shown in Figure \ref{fig:gpuusage}, the NRP implemented an allocation system aimed at optimizing GPU utilization. Following the introduction of the reservation system for A100 GPUs, average GPU utilization across the cluster improved significantly from 21.49\% to 31.37\%. Although there is still room for further enhancement, this marked improvement underscores the effectiveness of the reservation system.

In addition to traditional NRP workloads submitted via Kubernetes or managed through Jupyter notebooks, the NRP cluster is also utilized by low-priority, preemptible jobs sourced from the OSG. This strategy maintains consistent cluster activity by running tasks that can be easily interrupted if higher-priority workloads arise, simultaneously offering substantial resources to the OSG community. In 2024 alone, the NRP contributed 6.5 million CPU hours to OSG projects.


\subsection{Storage}

The National Research Platform employs diverse storage technologies, primarily relying on Ceph\cite{weil2006ceph}, a distributed filesystem orchestrated by the Rook operator, to deliver substantial storage capacity. Users access Ceph storage via Kubernetes Persistent Volumes. They can choose CephFS, a shared filesystem with managed namespaces, or Ceph Block Devices, ideal for handling smaller files. The distributed nature of NRP poses both challenges and opportunities. Among the challenges is the inherent data access latencies in a globally distributed system. To ensure minimal latency and optimal performance for users across the globe, Ceph storage is strategically distributed across five U.S. regions accessible to general users: West Coast, Central (Great Plains), East Coast, Southeast (Florida), and Pacific (including Guam and Hawaii). Users select storage regions, aligning computational workloads to the same region, minimizing latency and optimizing performance.

The NRP minimizes the impact of networking, power, or other physical failures in one location by distributing storage across multiple locations, and configuring each regional Ceph cluster to distribute data replicas across multiple physical locations. Any one location may experience an outage without leading to any filesystem outage. This strategy increases resilience while distributing the network and IO load across a region. 



\section{Security}
\label{sec:security}

One significant challenge for our project was effectively detecting and responding to malicious activities, particularly incidents involving cryptominers executed through unsecured user-operated services like JupyterHub instances. Our initial detection strategy relied on Prometheus alerts triggered by node overloads, which often represented legitimate usage, underscoring the need for explicit alerts pinpointing malicious activity. After evaluating several options, we selected Falco\cite{noauthor_falco_nodate}, an open-source threat detection tool sponsored by the Cloud Native Computing Federation (CNCF), widely adopted by the Kubernetes community.

Falco integrates with Kubernetes and container runtimes, capturing detailed process information (e.g., commands, arguments, hierarchies) along with metadata such as container images, namespaces, and pod identifiers. Its extensible architecture enables the implementation of custom and community-developed rules, significantly enhancing our ability to rapidly identify and respond to cryptomining and other malicious threats within the NRP environment.

\section{Accounting and Measurements}

Initially, we used standard Kubernetes tools (Prometheus, Thanos, Grafana) for accounting and monitoring. As data demands increased, Grafana became insufficient, prompting us to switch to ObservableHQ \cite{observablehq} for interactive, JavaScript-based dashboards. To overcome Thanos scaling limitations with accounting queries, we implemented custom caching, query segmentation, and a Redis caching layer – significantly improving performance and stability.

Beyond accounting, we maintain continuous network and system monitoring. We use sFlow for detailed network flow data, analyzed via ElastiFlow and InMon, and deploy distributed PerfSONAR instances across all nodes to consistently track network bandwidth and connectivity.

\section{Future Directions}

The default Kubernetes scheduler’s FIFO behavior limits the NRP’s capability to handle diverse scientific workloads optimally. To overcome this, we have begun experimenting with advanced scheduler plugins such as YuniKorn\cite{yunikorn}, which introduces sophisticated scheduling capabilities including quotas, reservations, and fair-share scheduling. Our ongoing collaboration with the YuniKorn development community aims to resolve identified issues and enhance scheduler performance.

The NRP is increasingly expanding to include smaller educational institutions, such as smaller R2 universities and community colleges like the San Diego Community College District. These institutions are not only contributing resources but also utilizing the NRP to support instructional programs in AI and machine learning, a trend we anticipate will continue to grow, broadening access and advancing workforce training across educational contexts.

\begin{acks}

This work was funded by the NSF awards 2112167, 2120019.  This paper was edited using LLMs hosted on the NRP \cite{noauthor_nrp-managed_nodate} as well as commercial LLMs.

\end{acks}

\bibliographystyle{ACM-Reference-Format}
\bibliography{sample-base}

\end{document}